\documentclass[letterpaper,journal,10pt]{IEEEtran}

\usepackage{cite}

\usepackage{graphicx}

\usepackage[cmex10]{amsmath}
\usepackage{amssymb}
\usepackage{fixltx2e}

\usepackage{multirow}
\usepackage{booktabs}
\usepackage{dcolumn}
\usepackage{ifpdf}

\usepackage{bm}
\usepackage[varg]{txfonts}
\let\mathbb=\varmathbb
\DeclareSymbolFont{letters}{OML}{ztmcm}{m}{it}

\usepackage{tikz,pgf}
\usepackage{pgfplots}
\usetikzlibrary{shapes,positioning,arrows,decorations.markings,fit,calc,patterns}

\usepackage{subfig}

\usepackage{algorithm}
\usepackage{algorithmicx}
\usepackage{algpseudocode}

\usepackage{listings}

\usepackage{color}
\newcommand{\fixme}[2]{\ifx&#2&{\color{red}#1}\else{\color{red}FIXME\{}#1{\color{red}\}}\footnote{{\color{red}#2}}\PackageWarning{Fixme}{#1: #2}\fi}

\title{Fast List Decoders for Polar Codes}

\author{Gabi~Sarkis, Pascal~Giard,~\IEEEmembership{Student~Member,~IEEE}, Alexander~Vardy,~\IEEEmembership{Fellow,~IEEE}, Claude~Thibeault,~\IEEEmembership{Senior~Member,~IEEE}, and Warren~J.~Gross,~\IEEEmembership{Senior~Member,~IEEE}%
\thanks{Manuscript received. The associate editor coordinating the review of this paper and approving it for publication was ...}%
\thanks{G. Sarkis, P. Giard, and W. J. Gross are with the Department of Electrical and Computer Engineering, McGill University, Montr\'eal, Qu\'ebec, Canada (e-mail: \{gabi.sarkis, pascal.giard\}@mail.mcgill.ca, warren.gross@mcgill.ca).}%
\thanks{A. Vardy is with the Department of Electrical Engineering, University of California San Diego, La Jolla, CA. USA (e-mail: avardy@ucsd.edu).}%
\thanks{C. Thibeault is with the Department of Electrical Engineering, \'Ecole de technologie sup\'erieure, Montr\'eal, Qu\'ebec, Canada (e-mail: claude.thibeault@etsmtl.ca).}}

\ifpdf
\fi

\DeclareMathOperator*{\argmin}{arg\,min}
\DeclareMathOperator*{\argmax}{arg\,max}

\newcommand{\est}[1]{\hat{u}_{#1}}

\newcommand{\estvec}[2]{\hat{\mvec{u}}_{#1}^{#2}}

\newcommand{\sgn}[1]{\text{sgn}(#1)}
\newcommand{\mvec}[1]{\bm{#1}}
\newcommand{\prob}[1]{\text{Pr}[#1]}
\newcommand{\tparam}[1]{\textless#1\textgreater}

\newcommand{\mb}[1]{\text{min}_{#1}}
\newcommand{\PM}[2]{\text{PM}_{#1}^{#2}}
\newcommand{\ma}[1]{|\alpha_v[\mb{#1}]|}
\newcommand{\latency}[1]{\mathcal{L}(#1)}

\begin{document}

\maketitle

\begin{abstract}
Polar codes asymptotically achieve the symmetric capacity of memoryless channels, yet their error-correcting performance under successive-cancellation (SC) decoding for short and moderate length codes is worse than that of other modern codes such as low-density parity-check (LDPC) codes. Of the many methods to improve the error-correction performance of polar codes, list decoding yields the best results, especially when the polar code is concatenated with a cyclic redundancy check (CRC). List decoding involves exploring several decoding paths with SC decoding, and therefore tends to be slower than SC decoding itself, by an order of magnitude in practical implementations.
In this paper, we present a new algorithm based on unrolling the decoding tree of the code that improves the speed of list decoding by an order of magnitude when implemented in software. Furthermore, we show that for software-defined radio applications, our proposed algorithm is faster than the fastest software implementations of LDPC decoders in the literature while offering comparable error-correction performance at similar or shorter code lengths.
\end{abstract}


\section{Introduction}
\label{sec:intro}
Polar codes, proposed by Ar\i{}kan \cite{Arikan2009}, achieve the symmetric capacity of memoryless channels as the code length $N \rightarrow \infty$ using the low-complexity successive-cancellation (SC) decoding algorithm. Their error-correction performance, however, is mediocre for codes of short and moderate lengths (a few thousand bits) and is worse than that of other modern codes, such as low-density parity-check (LDPC) codes. To improve their performance, polar codes are concatenated with a cyclic redundancy check (CRC) as an outer code and decoded using the list decoding algorithm (``list-CRC''). The resulting error-correction performance can exceed that of LDPC codes of similar length \cite{Tal2015}.

However, list-CRC decoding comes with a downside: the sequential ``bit-by-bit'' decoding order of the SC algorithm limits the speed of practical implementations, which further decreases with increasing list size $L$. The complexity of SC decoding is $O(N \log N)$, however a list decoder has a higher complexity of $O(L N \log N)$. The result is that practical hardware and software implementations of list decoders have low throughput that is an order of magnitude lower than the fastest SC decoder hardware \cite{Sarkis2014},  which achieves information throughout of $1.0$ Gbps at $100$ MHz in FPGA. The fastest belief propagation polar decoder is also faster: it achieves $2.34$ Gbps at $300$ MHz in 65nm CMOS \cite{Park2014}. On the other hand, reported hardware list decoder implementations achieve coded throughputs of 285 Mbps at 714 MHz for $N = 1024$ and $L = 2$ \cite{Lin2014}, and 335 Mbps at 847 MHz for $N = 1024$ and $L = 2$ \cite{BalatsoukasStimming2015}. For a list size $L=16$, the fastest decoder has a coded throughput of 220 Mbps at a clock frequency of 641 MHz \cite{Fan2015}.

The key to increasing the speed of SC decoders is to break the serial constraint imposed by successive cancellation. In \cite{Alamdar-Yazdi2011}, it was recognized that certain decoding steps in SC decoding were redundant for certain groups of bits that could instead be estimated simultaneously, given appropriate implementations. 
In that approach, called \emph{simplified successive cancellation} (SSC), groups of frozen bits do not need to be explicitly decoded, since their values are already known (usually zero), and groups of information bits can be estimated by thresholding, instead of serial successive cancellation. When viewing the polar code in a tree representation, it is easy to see that the code is a concatenation of smaller constituent codes. Groups of frozen bits can be viewed as comprising a ``Rate-0'' code and information bit groups are a ``Rate-1'' code.
Later work further increased the speed of SC decoding by parallel decoding some of the other ``Rate-$R$'' codes in the tree \cite{Sarkis2013, Sarkis2014}. The Fast-SSC algorithm in \cite{Sarkis2014} considers a variety of different constituent codes, such as single-parity-check (SPC) and repetition codes, decoding them with parallel hardware, estimating several bits per clock cycle. The first portion of this work describes how the Fast-SSC decoding algorithm was adapted for use in the context of list decoding.

The second part describes how this algorithm performs when implemented on a general purpose processor using single-instruction multiple-data (SIMD) instructions. Such systems were shown to have fast software SC decoders: the decoder in \cite{LeGal2014} employs inter-frame parallelism, decoding many frames in parallel, to achieve information throughput of 2.2 Gbps and latency of 26 $\mu$s. Alternatively, intra-frame parallelism targeting low-latency implementations was used by \cite{Giard2015} to reach an information throughput up to 1.3 Gbps with 1 $\mu$s latency. In addition, encoding of polar codes is a low complexity, $O(N \log N)$, operation that is well suited for software implementation as it does not require permutation of data \cite{Sarkis_TCOM_2015}.

The low encoding complexity combined with the good error-correction performance of list-CRC decoding will significantly improve the communication ability of wireless sensors networks (WSN) using software-defined radio (SDR). The sensor nodes benefit from the ability to use shorter codes, reducing transmission time and energy as well as the ability to reduce transmission power. Alternatively, instead of reducing transmission power, one can increase the distance between the nodes and base stations, reducing the number of base stations in the process. The nodes also benefit from the very low complexity of polar encoders \cite{Yoo2015,Sarkis_TCOM_2015}. The base stations, which generally have less stringent energy requirements, can use general purpose processors, including SIMD capable embedded ARM processors, to implement the proposed list-CRC decoding algorithm and to process data on site. This reduces the cost and development time of the WSN and increases its flexibility as a result of the SDR components.
This work could also be used in other SDR applications that do not have the  scale to justify a custom hardware implementation but where a throughput in the tens of Mbps is desirable. Quantum key distribution is such an example where a general purpose processor or a graphics processing unit is used to perform error correction \cite{Jouguet2014}. Multiple SDR systems either include a general purpose Intel processor or must be connected to a computer \cite{NutaqPicoSDR,EttusUSRPN,EttusUSRPB}, providing target platforms where our proposed algorithm can be used.

This work expands and improves on previous conference publications \cite{Sarkis2014b} and \cite{Sarkis2014a}. The algorithm in this paper has been reformulated in terms of log-likelihood ratios (LLRs), which yields speed improvements over the preliminary work in \cite{Sarkis2014b}. Furthermore, the conference paper implemented a list decoding algorithm based on SSC decoding (list-SSC), while this work develops the Fast-SSC algorithm for list decoding (list-Fast-SSC) and implements it, yielding further performance improvements. In addition, a general path metric is derived from codeword likelihoods, which is then used as the basis for calculating all the specialized decoders' output metrics.
Finally, unrolling \cite{Sarkis2014a} is applied to list decoders in this work. The results show that our improved list decoding algorithm results in a speedup of $11.9$ times compared to LLR-based list-SC decoding.
In addition to the decoder in \cite{Sarkis2014b}, those of \cite{Xiong2015} and \cite{Yuan2014} also perform multi-bit decisions and decoding. The main difference between them and the proposed decoder is that the former perform multi-bit decision for any constituent code of length $M$ bits using an exhaustive-search decoding algorithm. Whereas the proposed decoder uses specialized, low-complexity algorithms to decode any constituent code to which these algorithms apply, regardless of the constituent code length. A version of \cite{Yuan2014} limited to 2-bit constituent codes appears in \cite{Yuan2015}.

It should be noted that multi-bit decoding for Rate-1, Rate-0, and repetition constituent codes was proposed in the context of likelihood-based Reed-Muller (RM) decoders \cite{Dumer2006} and likelihood-based RM list decoders \cite{Dumer2006a}. This work differs from \cite{Dumer2006a} in that it targets LLR-based list decoders in the context of polar codes and recognizes more special constituent multi-bit decoders. The algorithms introduced in this work focus on low implementation complexity, especially for SIMD processors and parallel hardware.

This work starts by reviewing the construction of polar codes and the list-CRC and the Fast-SSC decoding algorithms in Section~\ref{sec:bg}. We then describe how to generate a software polar decoder amenable to vectorization in Section~\ref{sec:unrolled}. Section~\ref{sec:algo} introduces the proposed list decoding algorithm and a software implementation is described in Section~\ref{sec:impl}. The speed and error-correction performance of the proposed decoder are studied in Section~\ref{sec:perf} and compared to those of LDPC codes of the 802.3an \cite{802.3an} and 802.11n \cite{802.11n} standards in Section~\ref{sec:ldpc}. In the second comparison, we show that polar codes can match or exceed the speed and error-correction performance of software LDPC decoders while using shorter codes.

\section{Background}
\label{sec:bg}

\subsection{Polar Codes}
\label{sec:bg:polar-codes}
A polar code of length $N$ is constructed recursively from two polar codes of length $N/2$. Successive-cancellation (SC) decoding provides a bit estimate $\hat{u}_i$ using the channel output $y_0^{N-1}$ and the previously estimated bits $\hat{u}_0^{i - 1}$ according to
\begin{equation}
\label{eq:sc}
\est{i} = \begin{cases}
  0 & \text{when } \prob{\mvec{y}, \estvec{0}{i-1} | \est{i} = 0} \geq \prob{\mvec{y}, \estvec{0}{i-1} | \est{i} = 1};\\
  1 & \text{otherwise}.
\end{cases}
\end{equation}

As $N \to \infty$, the probability of correctly estimating a bit approaches 1 or 0.5. This is the channel polarization phenomenon that is exploited by polar codes, which use reliable bit locations to store information bits and set the unreliable, called frozen, bits to zero. As a result, when the SC decoder is estimating a bit $u_i$, it is zero if the bit is frozen, or is calculated according to \eqref{eq:sc}.

\begin{figure}[t]
  \centering
  \subfloat[Graph]{\label{fig:sc-graph}\newcommand{\ubit}[1]{$u_{#1}$}
\newcommand{\fbit}[1]{\color{gray}$u_{#1}$}
\begin{tikzpicture}[baseline=(s37.center)]

\usetikzlibrary{shapes,positioning,arrows,decorations.markings,fit}

\definecolor{varnode_fill}{RGB}{0,0,0}
\definecolor{chknode_fill}{RGB}{255,255,255}

\tikzset{
  chknode/.style={draw,fill=chknode_fill,circle,minimum size=0.3cm, inner sep=0},
  varnode/.style={draw,fill=varnode_fill,circle,minimum size=0.1cm, inner sep=0},
  sep/.style={rectangle,minimum width=0.25cm, inner sep=0},
  bit/.style={circle, inner sep = 0}
}

\tikzset{blue dotted/.style={draw=blue!50!white, line width=1pt,
    dash pattern=on 4pt off 4pt,
    inner sep=0.5mm, rectangle, rounded corners}};

\tikzset{blue dotted tight/.style={draw=blue!50!white, line width=1pt,
    dash pattern=on 4pt off 4pt,
    inner sep=0mm, rectangle, rounded corners}};

\matrix[row sep=1mm, column sep=1mm] {
	\node[bit] (n0s0) {\fbit{0}}; & \node[sep] (s0) {}; & \node[chknode] (n0s1) {$+$}; & \node[sep] (s10) {}; &&	\node[chknode] (n0s2) {$+$}; &	\node[sep] (s20) {}; &&&& \node[chknode] (n0s3) {$+$}; &    \node[sep] (s30) {}; & \node[circle] (n0s4) {}; \\ 
	\node[bit] (n1s0) {\fbit{1}}; & \node[sep] (s1) {}; & \node[varnode] (n1s1) {};	   & \node[sep] (s11) {}; &	 \node[chknode] (n1s2) {$+$}; && \node[sep] (s21) {}; &&&  \node[chknode] (n1s3) {$+$}; &&   \node[sep] (s31) {}; & \node[circle] (n1s4) {}; \\ 
	\node[bit] (n2s0) {\fbit{2}}; & \node[sep] (s2) {}; & \node[chknode] (n2s1) {$+$}; & \node[sep] (s12) {}; &&	 \node[varnode] (n2s2) {};    &	 \node[sep] (s22) {}; &&   \node[chknode] (n2s3) {$+$}; &&&  \node[sep] (s32) {}; & \node[circle] (n2s4) {}; \\ 
	\node[bit] (n3s0) {\ubit{3}}; & \node[sep] (s3) {}; & \node[varnode] (n3s1) {};	   & \node[sep] (s13) {}; &	 \node[varnode] (n3s2) {};    && \node[sep] (s23) {}; &	   \node[chknode] (n3s3) {$+$}; &&&& \node[sep] (s33) {}; & \node[circle] (n3s4) {}; \\ 
	\node[bit] (n4s0) {\fbit{4}}; & \node[sep] (s4) {}; & \node[chknode] (n4s1) {$+$}; & \node[sep] (s14) {}; &&	 \node[chknode] (n4s2) {$+$}; &	 \node[sep] (s24) {}; &&&& \node[varnode] (n4s3) {};	&    \node[sep] (s34) {}; & \node[circle] (n4s4) {}; \\ 
	\node[bit] (n5s0) {\ubit{5}}; & \node[sep] (s5) {}; & \node[varnode] (n5s1) {};	   & \node[sep] (s15) {}; &	 \node[chknode] (n5s2) {$+$}; && \node[sep] (s25) {}; &&&  \node[varnode] (n5s3) {};	&&   \node[sep] (s35) {}; & \node[circle] (n5s4) {}; \\ 
	\node[bit] (n6s0) {\ubit{6}}; & \node[sep] (s6) {}; & \node[chknode] (n6s1) {$+$}; & \node[sep] (s16) {}; &&	 \node[varnode] (n6s2) {};    &	 \node[sep] (s26) {}; &&   \node[varnode] (n6s3) {};	&&&  \node[sep] (s36) {}; & \node[circle] (n6s4) {}; \\ 
	\node[bit] (n7s0) {\ubit{7}}; & \node[sep] (s7) {}; & \node[varnode] (n7s1) {};	   & \node[sep] (s17) {}; &	 \node[varnode] (n7s2) {};    && \node[sep] (s27) {}; &	   \node[varnode] (n7s3) {};	&&&& \node[sep] (s37) {}; & \node[circle] (n7s4) {}; \\ 
};
\path[-] (n0s0) edge (n0s1) (n0s1) edge (n0s2) (n0s2) edge (n0s3) (n0s3) edge (n0s4);
\path[-] (n1s0) edge (n1s1) (n1s1) edge (n1s2) (n1s2) edge (n1s3) (n1s3) edge (n1s4);
\path[-] (n2s0) edge (n2s1) (n2s1) edge (n2s2) (n2s2) edge (n2s3) (n2s3) edge (n2s4);
\path[-] (n3s0) edge (n3s1) (n3s1) edge (n3s2) (n3s2) edge (n3s3) (n3s3) edge (n3s4);
\path[-] (n4s0) edge (n4s1) (n4s1) edge (n4s2) (n4s2) edge (n4s3) (n4s3) edge (n4s4);
\path[-] (n5s0) edge (n5s1) (n5s1) edge (n5s2) (n5s2) edge (n5s3) (n5s3) edge (n5s4);
\path[-] (n6s0) edge (n6s1) (n6s1) edge (n6s2) (n6s2) edge (n6s3) (n6s3) edge (n6s4);
\path[-] (n7s0) edge (n7s1) (n7s1) edge (n7s2) (n7s2) edge (n7s3) (n7s3) edge (n7s4);

\path[-] (n0s1) edge (n1s1);
\path[-] (n2s1) edge (n3s1);
\path[-] (n4s1) edge (n5s1);
\path[-] (n6s1) edge (n7s1);

\path[-] (n0s2) edge (n2s2);
\path[-] (n1s2) edge (n3s2);
\path[-] (n4s2) edge (n6s2);
\path[-] (n5s2) edge (n7s2);

\path[-] (n0s3) edge (n4s3);
\path[-] (n1s3) edge (n5s3);
\path[-] (n2s3) edge (n6s3);
\path[-] (n3s3) edge (n7s3);

\node (g_n0s0) [blue dotted tight, fit = (n0s0)] {};
\node (g_n1s0) [blue dotted tight, fit = (n1s0)] {};
\node (g_n2s0) [blue dotted tight, fit = (n2s0)] {};
\node (g_n3s0) [blue dotted tight, fit = (n3s0)] {};
\node (g_n4s0) [blue dotted tight, fit = (n4s0)] {};
\node (g_n5s0) [blue dotted tight, fit = (n5s0)] {};
\node (g_n6s0) [blue dotted tight, fit = (n6s0)] {};
\node (g_n7s0) [blue dotted tight, fit = (n7s0)] {};

\node (g_n0s1) [blue dotted, fit = (n0s1) (n1s1)] {};
\node (g_n1s1) [blue dotted, fit = (n2s1) (n3s1)] {};
\node (g_n2s1) [blue dotted, fit = (n4s1) (n5s1)] {};
\node (g_n3s1) [blue dotted, fit = (n6s1) (n7s1)] {};

\node (g_n0s2) [blue dotted, fit = (n0s2) (n1s2) (n2s2) (n3s2)] {};
\node (g_n1s2) [blue dotted, fit = (n4s2) (n5s2) (n6s2) (n7s2)] {};

\node (g_n0s3) [blue dotted, fit = (n0s3) (n1s3) (n2s3) (n3s3) (n4s3) (n5s3) (n6s3) (n7s3)] {};

\end{tikzpicture}}
  \quad
  \subfloat[Decoder tree]{\label{fig:sc-tree} \begin{tikzpicture}[baseline = (0_7.center),
        level/.style={level distance = 6mm},
        level 1/.style={sibling distance=19mm, edge from parent/.style={draw,black,line width=2pt}},
        level 2/.style={level distance=10mm, sibling distance=9.5mm, edge from parent/.style={draw,black,line width=1pt}},
        level 3/.style={sibling distance=4.7mm, edge from parent/.style={draw,black,line width=0.5pt}},
        ]

\tikzset{
frozen/.style={thick,draw=black,fill=white,minimum size=3mm,circle, inner sep=0},
fullspace/.style={thick,draw=black,fill=black,minimum size=3mm,circle, inner sep = 0},
mixed/.style={thick,draw=black,fill=gray,minimum size=3mm,circle, inner sep = 0},
ml_mixed/.style={thick,draw=black,fill=blue,minimum size=3mm,circle, inner sep = 0}
}

\node[mixed] (p){} [grow=left]
	child {node[mixed] (2_0){}
		child {node[mixed] (1_0){}
			child {node[frozen] (a0_0){}
			}
			child {node[frozen] (a0_1){}
			}
		}
		child {node[mixed] (1_2){}
			child {node[frozen] (0_2){}
			}
			child {node[fullspace] (0_3){}
			}
		}
	}
	child {node[mixed] (v){$v$}
		child {node[mixed, label={[label distance = -1mm]left}] (cl){}
			child {node[frozen] (0_4){}
			}
			child {node[fullspace] (0_5){}
			}
		}
		child {node[mixed, label={[label distance = -1mm]below:right}] (cr){}
			child {node[fullspace] (0_6){}
			}
			child {node[fullspace] (0_7){}
			}
		}
	}
;

\draw[<-] ($(v.north east) - (1mm, -1mm)$) -- node[above left=-1.5mm] {\footnotesize $\alpha_v$} ($(p.south west) - (1mm, 0mm)$);
\draw[<-] ($(p.south west) + (1mm, -1mm)$) -- node[below right=-2mm] {\footnotesize $\beta_v$} ($(v.north east) + (1mm, 0mm)$);

\draw[<-] ($(v.north west) + (-1.6mm, -1mm)$) -- node[left=1mm] {\footnotesize $\beta_l$} ($(cl.south east) + (0mm, -0.6mm)$);
\draw[<-] ($(cl.south east) - (-1.6mm, -1mm)$) -- node[above right=-2mm] {\footnotesize$\alpha_l$} ($(v.north west) - (0mm, -0.6mm)$);

\draw[<-] ($(v.south west) + (0mm, -0.6mm)$) -- node[below right=-2mm] {\footnotesize $\beta_r$} ($(cr.north east) + (1.6mm, -1mm)$);
\draw[<-] ($(cr.north east) - (0mm, -0.6mm)$) -- node[left=1mm] {\footnotesize $\alpha_r$} ($(v.south west) - (1.6mm, -1mm)$);

\end{tikzpicture}}
  \caption{The graph of an (8, 4) polar code and its corresponding tree representation.}
  \label{fig:graph-to-tree}
\end{figure}
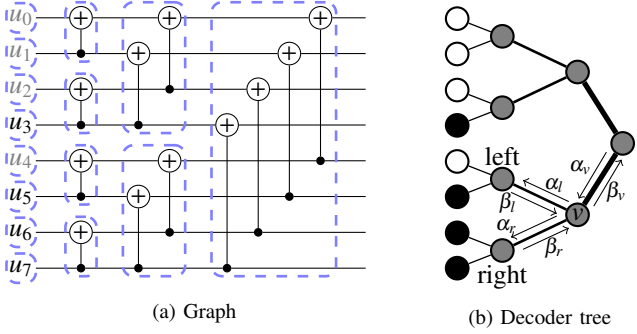

Fig.~\ref{fig:sc-graph} shows the graph of an (8, 4) polar code where frozen bits are labeled in gray and information bits in black. The SC decoder can also be viewed as a tree that is traversed depth first. Such a tree is illustrated in Fig.~\ref{fig:sc-tree}, where each sub-tree corresponds to a constituent code. The white nodes correspond to frozen bits, and the black ones to information bits. The gray nodes represent the concatenation operations combining two constituent codes.

Two types of messages are passed along the edges of the tree in the decoder: soft reliability values---LLRs in this work,---$\alpha$, and hard bit estimates, $\beta$. When a node corresponding to a constituent code of length $N_v$ receives the reliability values from its parent, represented using LLRs, the output to its left child is calculated according to the $F$ function:
\begin{align}
\label{eq:f}
\alpha_l[i] &= F (\alpha_v[i], \alpha_v[i + N_v/2] )\nonumber\\
  &= 2\text{atanh}(\tanh(\alpha_v[i]/2) \tanh(\alpha_v[i + N_v/2]/2))\nonumber\\
  &\approx \sgn{\alpha_v[i]} \sgn{\alpha_v[i + N_v/2]} \min ( |\alpha_v[i] |,|\alpha_v[i + N_v/2] |);
\end{align}
where the approximation is the min-sum approximation.

Once the output of the left child $\beta_l$ is available the message to right one is calculated using the $G$ function
\begin{align}
\label{eq:g}
\alpha_r[i] &= G(\alpha_v[i], \alpha_v[i + N_v/2], \beta_l[i])\nonumber\\
            &= \alpha_v[i + N_v/2] - (2 \beta_l[i] - 1) \alpha_v[i].
\end{align}
Finally, when $\beta_r$ is known, the node's output is computed as
\begin{equation}
\label{eq:combine}
\beta_v[i] = \begin{cases}
  \beta_l[i] \oplus \beta_r[i] & \text{when } i < N_v/2;\\
  \beta_r[i - N_v/2] & \text{otherwise;}
\end{cases}
\end{equation}
where $\oplus$ is an XOR operation and we refer to the operation as the \textit{Combine} operation.

The output $\beta_v$ of a frozen node is always zero, and is calculated using threshold detection for an information node:
\begin{equation}
\beta_v = h(\alpha_v) = \begin{cases}
  0 & \text{when } \alpha_v \geq 0;\\
  1 & \text{otherwise}.
\label{eq:sc:h}
\end{cases}
\end{equation}

\subsection{List-CRC Decoding}
\label{sec:bg:list-crc}
When estimating an information bit, a list decoder continues decoding along two paths, the first assumes that `0' was the correct bit estimate, and the second `1'. Therefore at every information bit, the decoder doubles the number of possible outcomes up to a predetermined limit $L$. When the number of paths exceeds $L$, the list is pruned by retaining only the $L$ most reliable paths. When decoding is over, the estimated codeword with the largest reliability metric is selected as the decoder output. It was observed in \cite{Tal2015} that using a CRC as the primary criterion for selecting the final decoder output, increased the error-correction performance significantly. In addition, the CRC enables the use of a adaptive decoder where the list size starts at two and is gradually increased until the CRC is satisfied or a maximum list size is reached \cite{Li2012}.

Initially, polar list decoders used likelihood \cite{Tal2015} and log-likelihood values \cite{BalatsoukasStimming2014a} to represent reliabilities. Later, log-likelihood ratios (LLRs) were used in \cite{BalatsoukasStimming2015} to reduce the amount of memory used by a factor of two and to reduce the processing complexity. In addition to the messages and operations presented in Section~\ref{sec:bg:polar-codes}, the algorithm of \cite{BalatsoukasStimming2015} stores a reliability metric $\text{PM}_l^i$ for each path $l$ that is updated for every estimated bit $i$ according to:
\begin{equation}
\text{PM}_l^i = \begin{cases}
  \text{PM}_l^{i - 1} & \text{if } \hat{u}_i = h(\alpha_v),\\
  \text{PM}_l^{i - 1}  - |\alpha_v| & \text{otherwise}.
\end{cases}
\label{eq:sc-list:info}
\end{equation}
It is important to note that the path metric is updated when encountering both information and frozen bits.

\subsection{Fast-SSC Decoding}
\label{sec:bg:fast-ssc}
The SC decoder traverses the code tree until reaching leaf nodes corresponding to codes of length one before estimating a bit. This was found to be superfluous as the output of sub-trees corresponding to constituent codes of rate 0 or rate 1 of any length can be estimated without traversing their sub-trees \cite{Alamdar-Yazdi2011}. The output of a rate-0 node is known a priori to be an all-zero vector of length $N_v$; while that of rate-1 can be found by applying threshold detection element-wise on $\alpha_v$ so that
\begin{equation}
\label{eq:info}
\beta_v[i] = h(\alpha_v[i]) = \begin{cases}
  0 & \text{when } \alpha_v[i] \geq 0;\\
  1 & \text{otherwise}.
\end{cases}
\end{equation}

The Fast-SSC algorithm utilizes low-complexity maximum-likelihood (ML) decoding algorithms to decode constituent repetition and single-parity check (SPC) codes instead of traversing their corresponding sub-trees \cite{Sarkis2013, Sarkis2014}.

The ML-decision for a repetition code is
\begin{equation}
\label{eq:rep}
\beta_v[i] = \begin{cases}
  0 & \text{when } \sum_j \alpha_v[j] \geq 0;\\
  1 & \text{otherwise}.
\end{cases}
\end{equation}

The SPC decoder performs threshold detection \eqref{eq:info} on its output to calculate the intermediate value \text{HD}. The parity of HD is calculated using modulo-2 addition and the least reliable bit is found according to
\[
j = \argmin_j |\alpha_v[j]|.
\]
The final output of the SPC decoder is
\begin{equation}
\label{eq:spc}
\beta_v[i] = \begin{cases}
  \text{HD}[i] \oplus \text{parity} & \text{when } i = j;\\
  \text{HD}[i] & \text{otherwise.}
\end{cases}
\end{equation}

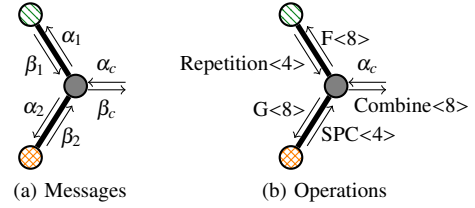
\begin{figure}[t]
  \centering
  \subfloat[Messages]{\label{fig:fast-ssc-tree}\definecolor{deepgreen}{RGB}{8, 130, 25}

\begin{tikzpicture}[baseline=(base),
        level/.style={level distance = 6mm},
        level 1/.style={sibling distance=19mm, edge from parent/.style={draw,black,line width=2pt}},
        level 2/.style={sibling distance=9mm, edge from parent/.style={draw,black,line width=1pt}},
        level 3/.style={sibling distance=4mm, edge from parent/.style={draw,black,line width=0.5pt}},
        ]

\tikzset{
frozen/.style={thick,draw=black,fill=white,minimum size=3mm,circle, inner sep=0},
fullspace/.style={thick,draw=black,fill=black,minimum size=3mm,circle, inner sep = 0},
mixed/.style={thick,draw=black,fill=gray,minimum size=3mm,circle, inner sep = 0},
rep_mixed/.style={thick,draw=black,pattern=north west lines,pattern color=deepgreen,minimum size=3mm,circle, inner sep = 0},
spc_mixed/.style={thick,draw=black,pattern=crosshatch,pattern color=orange,minimum size=3mm,circle, inner sep = 0},
repspc/.style={thick,draw=black,pattern=vertical lines,pattern color=blue,minimum size=3mm,circle, inner sep = 0}
}

\tikzset{
parallel segment/.style={
   segment distance/.store in=\segDistance,
   segment pos/.store in=\segPos,
   segment length/.store in=\segLength,
   to path={
   ($(\tikztostart)!\segPos!(\tikztotarget)!\segLength/2!(\tikztostart)!\segDistance!90:(\tikztotarget)$) -- 
   ($(\tikztostart)!\segPos!(\tikztotarget)!\segLength/2!(\tikztotarget)!\segDistance!-90:(\tikztostart)$)  \tikztonodes
   }, 
   segment pos=.5,
   segment length=4ex,
   segment distance=-1mm,
},
}

\node[mixed] (3_0){} [grow=left]
	child {node[rep_mixed] (2_0){}
	}
	child {node[spc_mixed] (2_1){}
	}
;
	
\draw[->] (3_0) to[parallel segment] node[above right=-1.5mm] {\footnotesize $\alpha_1$} (2_0);
\draw[->] (2_0) to[parallel segment] node[below left=-1.5mm] {\footnotesize $\beta_1$} (3_0);

\draw[->] (3_0) to[parallel segment] node[above left=-1.5mm] {\footnotesize $\alpha_2$} (2_1);
\draw[->] (2_1) to[parallel segment] node[below right=-1.5mm] {\footnotesize $\beta_2$} (3_0);

\draw[<-] ($(3_0.east) + (0mm, 0.5mm)$) -- node[above=0mm] {\footnotesize $\alpha_c$} ($(3_0.east) + (5mm, 0.5mm)$);
\draw[->] ($(3_0.east) + (0mm, -0.5mm)$) -- node[below=0mm] {\footnotesize $\beta_c$} ($(3_0.east) + (5mm, -0.5mm)$);

\node [circle, below= -5.27mm of 2_1.base] (base) {};
\end{tikzpicture}}
  \quad
  \subfloat[Operations]{\label{fig:fast-ssc-functions}\definecolor{deepgreen}{RGB}{8, 130, 25}

\begin{tikzpicture}[baseline=(base),
        level/.style={level distance = 6mm},
        level 1/.style={sibling distance=19mm, edge from parent/.style={draw,black,line width=2pt}},
        level 2/.style={sibling distance=9mm, edge from parent/.style={draw,black,line width=1pt}},
        level 3/.style={sibling distance=4mm, edge from parent/.style={draw,black,line width=0.5pt}},
        ]

\tikzset{
frozen/.style={thick,draw=black,fill=white,minimum size=3mm,circle, inner sep=0},
fullspace/.style={thick,draw=black,fill=black,minimum size=3mm,circle, inner sep = 0},
mixed/.style={thick,draw=black,fill=gray,minimum size=3mm,circle, inner sep = 0},
rep_mixed/.style={thick,draw=black,pattern=north west lines,pattern color=deepgreen,minimum size=3mm,circle, inner sep = 0},
spc_mixed/.style={thick,draw=black,pattern=crosshatch,pattern color=orange,minimum size=3mm,circle, inner sep = 0},
repspc/.style={thick,draw=black,pattern=vertical lines,pattern color=blue,minimum size=3mm,circle, inner sep = 0}
}

\tikzset{
parallel segment/.style={
   segment distance/.store in=\segDistance,
   segment pos/.store in=\segPos,
   segment length/.store in=\segLength,
   to path={
   ($(\tikztostart)!\segPos!(\tikztotarget)!\segLength/2!(\tikztostart)!\segDistance!90:(\tikztotarget)$) -- 
   ($(\tikztostart)!\segPos!(\tikztotarget)!\segLength/2!(\tikztotarget)!\segDistance!-90:(\tikztostart)$)  \tikztonodes
   }, 
   segment pos=.5,
   segment length=4ex,
   segment distance=-1mm,
},
}

\node[mixed] (3_0){} [grow=left]
	child {node[rep_mixed] (2_0){}
	}
	child {node[spc_mixed] (2_1){}
	}
;
	
\draw[->] (3_0) to[parallel segment] node[above right=-1.5mm] {\footnotesize F\tparam{8}} (2_0);
\draw[->] (2_0) to[parallel segment] node[below left=-1.5mm] {\footnotesize Repetition\tparam{4}} (3_0);

\draw[->] (3_0) to[parallel segment] node[above left=-1.5mm] {\footnotesize G\tparam{8}} (2_1);
\draw[->] (2_1) to[parallel segment] node[below right=-1.5mm] {\footnotesize SPC\tparam{4}} (3_0);

\draw[<-] ($(3_0.east) + (0mm, 0.5mm)$) -- node[above=0mm] {\footnotesize $\alpha_c$} ($(3_0.east) + (5mm, 0.5mm)$);
\draw[->] ($(3_0.east) + (0mm, -0.5mm)$) -- node[label={[label distance=-3mm]-2mm:{\footnotesize Combine\tparam{8}}}] {} ($(3_0.east) + (5mm, -0.5mm)$);

\end{tikzpicture}}
  \caption{Fast-SSC decoder graph for an $(8,4)$ polar code.}
  \label{fig:fast-ssc}
\end{figure}

Fig.~\ref{fig:fast-ssc} shows a Fast-SSC decoder tree for the (8, 4) code, indicating the messages passed in the decoder and the operations used to calculate them.

The Fast-SSC decoder and its software implementation \cite{Giard2015} utilize additional specialized constituent decoders that are not used in this work as they did not improve decoding speed. In addition, the operations mentioned in this section and implemented in \cite{Giard2015} present a single output and therefore cannot be applied directly to list decoding. In this work, we will show how they are adapted to present multiple candidates and used in a list decoder.

\subsection{Unrolling Software Decoders}
\label{sec:unrolled}
The software list decoder in \cite{Sarkis2014b} is run-time configurable, i.e. the same executable is capable of decoding any polar code without recompilation. While flexible, this limits the achievable decoding speed. In \cite{Sarkis2014a}, it was shown that generating a decoder for a specific polar code yielded significant speed improvement by replacing branches with straight-line code and increasing the utilization of SIMD instructions. This process is managed by a developed CAD tool that divides the process into two parts: decoder tree optimization, and C++ code generation.

For the list decoder in this paper we applied this optimization tool using a subset of the nodes available to the complete Fast-SSC algorithm: Rate-0 (Frozen), Rate-1 (information), repetition, and SPC nodes. The decoder tree optimizer traverses the decoder tree starting from its root. If a sub-tree rooted at the current node has a higher decoding latency than an applicable Fast-SSC node, it is replaced with the latter. If there are not any Fast-SSC nodes that can replace the current tree, the optimizer moves to the current node's children and repeats the process.

Once the tree is optimized, the corresponding C++ code is generated. All functions are passed the current $N_v$ value as a template parameter, enabling vectorization and loop unrolling. 

Listings \ref{lst:loop-decoder} and \ref{lst:fast-ssc} show a loop-based decoder and an unrolled one for the (8, 4) code in Fig.~\ref{fig:fast-ssc}, respectively. In the loop-based decoder, both iterating over the decoding operations and selecting an appropriate decoding function (called an operation processor) to execute involve branches. In addition, the operation processor does not know the size of the data it is operating on at compile-time; and as such, it must have another loop inside. The unrolled decoder can eliminate these branches since both the decoder flow and data sizes are known at compile-time.

{
\floatname{algorithm}{Listing}
\begin{algorithm}[t]
\caption{Loop-based (8, 4) Fast-SSC Decoder}
\label{lst:loop-decoder}
\begin{lstlisting}[mathescape]
for (unsigned int i = 0; i < operation_count; ++i) {
  operation_processor = fetch_operation_processor(i);
  operation_processor.execute($\alpha_v$, &$\alpha_r$, &$\alpha_l$, $\beta_l$, $\beta_r$, &$\beta_v$);
}
\end{lstlisting}
\end{algorithm}

\begin{algorithm}[t]
\caption{Unrolled (8, 4) Fast-SSC Decoder}
\label{lst:fast-ssc}
\begin{algorithmic}
  \State $\alpha_1$ = $F$\tparam{8}($\alpha_c$);
  \State $\beta_1$ = Repetition\tparam{4}($\alpha_1$);
  \State $\alpha_2$ = $G$\tparam{8}($\alpha_c$, $\beta_1$);
  \State $\beta_2$ = SPC\tparam{4}($\alpha_2$);
  \State $\beta_c$ = \textit{Combine}\tparam{8}($\beta_1$, $\beta_2$);
\end{algorithmic}
\end{algorithm}
}
\section{Proposed List-Decoding Algorithm}
\label{sec:algo}
When performing operations corresponding to a rate-$R$ node, a list decoder with a maximum list size $L$ performs the operations $F$ \eqref{eq:f}, $G$ \eqref{eq:g}, and \textit{Combine} \eqref{eq:combine} on each of the paths independently. It is only at the leaf nodes that interaction between the paths occurs: the decoder generates new paths and retains the most reliable $L$ ones.

A significant difference between the baseline SC-list decoder and the proposed algorithm is that each path in the former generates two candidates, whereas in the latter, the leaf nodes with sizes larger than one can generate multiple candidates for each path.

All path-generating nodes store the candidate path reliability metrics in a priority queue so that the worst candidate can be quickly found and replaced with a new path when appropriate. This is an improvement over \cite{Sarkis2014b}, where path reliability metrics were kept sorted at all times by using a red-black (RB) tree. The most common operation in candidate selection is locating the path with the minimum reliability, which is an $O(\log L)$ operation in RB-trees, the order of the remaining candidates is irrelevant. A heap-backed priority queue provides $O(1)$ minimum-value look up and $O(\log L)$ insertion and removal, and is therefore more efficient than an RB tree for the intended application.

In this section, we describe how each node generates its output paths and calculates the corresponding reliability metrics. The process of retaining the $L$ most reliable paths is described in Algorithm~\ref{algo:l-candidates}. Performing the candidate selection in two passes and storing the ML decisions first are necessary to prevent candidates generated by the first few paths from overwriting the input for later ones.
\begin{algorithm}[t]

\caption{Candidate selection process}
\label{algo:l-candidates}
\begin{algorithmic}
\For {$s \in \textit{sourcePaths}$}
\State Generate candidates.
\State Store reliability of all candidates except the ML one.
\State Store ML decision.
\EndFor
\For {$p \in \textit{candidates}$}
\If {fewer than $L$ candidates are stored}
\State Store $p$.
\ElsIf {$\PM{p}{t} < $ min. stored candidate reliability}
\State Replace min. reliability candidate with $p$.
\EndIf
\EndFor
\end{algorithmic}
\end{algorithm}

\subsection{Candidate Generation and Reliability}
The aim of the proposed algorithm is to directly generate candidates without traversing sub-trees whenever possible. To achieve this goal, we use the candidate-enumeration method of Chase decoding \cite{Chase1972} to provide a list of candidate paths at the output of a rate-1 decoder.

The log-likelihood of a candidate codeword $\beta_j$ is
\begin{align}
l(\beta_j) &= \sum_i (1 - 2\beta_j[i]) \alpha_v[i] \nonumber\\
& = \sum_i (1 - 2\beta_j[i]) \sgn{\alpha_v[i]} \left|\alpha_v[i]\right| \nonumber\\
& = \sum_i (1 - 2\left|\beta_j[i] - h(\alpha_v[i])\right|) \left|\alpha_v[i]\right|
\end{align}

The factor
\begin{align}
(1 - 2\beta_j[i]) \sgn{\alpha_v[i]} &= 1 - 2\left|\beta_j[i] - h(\alpha_v[i])\right| \nonumber\\
&=\begin{cases}
+1 & \text{when } \beta_j[i] = h(\alpha_v[i]),\nonumber\\
-1 & \text{otherwise.}
\end{cases}
\end{align}

The ML candidate codeword is
\[
\beta_{\text{ML}} = \argmax_{\beta_i \in \mathcal{C}} l(\beta_i),
\]
where $\mathcal{C}$ is the set of all codewords. The other candidates are generated by flipping bits relative to the ML decision, both individually and simultaneously, subject to the constraint that the candidate is a valid codeword.

To ensure that the codeword log-likelihood values remain $\leq 0$, we offset $l(\beta_j)$ by $\sum_i |\alpha_v[i]|$. In addition, we scale the metric by a factor of $0.5$. The resulting codeword metric becomes
\begin{align}
l'(\beta_j) &= \frac{l(\beta_j) - \sum_i |\alpha_v[i]|}{2} \nonumber\\
&= \frac{\sum_i (1 - 2\left|\beta_j[i] - h(\alpha_v[i])\right|) \left|\alpha_v[i]\right| - \sum_i |\alpha_v[i]|}{2} \nonumber\\
&= \frac{\sum_i(1 - 2\left|\beta_j[i] - h(\alpha_v[i])\right| - 1)\left|\alpha_v[i]\right|}{2} \nonumber\\
&= -\sum_i \left|\beta_j[i] - h(\alpha_v[i])\right| \left|\alpha_v[i]\right|.
\end{align}
This metric states that a codeword is penalized for any difference between it and the vector calculated from $\alpha_v$ using \eqref{eq:info}.

When starting from a source path $s$ with reliability $\PM{s}{t-1}$, the reliability of the path corresponding to the codeword $\beta_j$ is
\begin{equation}
\label{eq:pm}
\PM{j}{t} = \PM{s}{t-1} -\sum_i \left|\beta_j[i] - h(\alpha_v[i])\right| \left|\alpha_v[i]\right|.
\end{equation}

All specialized decoders generate their candidates based on this metric by restricting potential codewords.

\subsection{Rate-0 Decoders}
\label{sec:algo:r0}

Rate-0 nodes do not generate new paths; however, like their length-1 counterparts in SC-list decoding \cite{BalatsoukasStimming2015}, they alter path reliability values. In \cite{BalatsoukasStimming2015}, the path metric was updated according to
\[
\text{PM}_l^i = \begin{cases}
  \text{PM}_l^{i - 1} & \text{if } h(\alpha_v) = 0,\\
  \text{PM}_l^{i - 1} - |\alpha_v| & \text{otherwise}.
\end{cases}
\]

The all-zero codeword, $\beta_j[i] = 0, \forall i$, is the only valid codeword. Therefore, based on \eqref{eq:pm}, the output path metric is
\begin{equation}
\label{eq:rel:frozen-h}
\PM{j}{t} = \PM{s}{t - 1} - \sum_i h(\alpha_v[i]) |\alpha_v[i]|.
\end{equation}

An alternative formulation for \eqref{eq:rel:frozen-h} is
\begin{equation}
\label{eq:rel:frozen}
\text{PM}_l^t = \text{PM}_l^{t - 1} - \sum_{i = 0}^{N_V -1} |\max(\alpha_v[i], 0)|.
\end{equation}

\subsection{Rate-1 Decoders}
\label{sec:algo:info}
A decoder for a length $N_v$ rate-1 constituent code can provide up to $2^{N_v}$ candidate codewords. This approach is impractical as it scales exponentially in $N_v$. The Chase-II decoding algorithm considers only a limited set of the least-reliable bits to generate its candidates \cite{Chase1972}. We use the same method to limit the complexity of rate-1 decoders when enumerating the candidates selected for consideration in \eqref{eq:pm}.

The maximum-likelihood decoding rule for a rate-1 code is \eqref{eq:info}. Additional candidates are generated by flipping the least reliable bits both independently and simultaneously. Empirically, we found that considering only the two least-reliable bits, whose indexes are denoted $\mb{1}$ and $\mb{2}$, is sufficient to match the performance of SC list decoding. Therefore, for each source path $s$, the proposed rate-1 decoder generates four candidates with the following reliability values
\begin{align}
\PM{0}{t} &= \PM{s}{t-1},\nonumber\\
\PM{1}{t} &= \PM{s}{t-1} - |\alpha_v[\mb{1}{}]|, \nonumber\\
\PM{2}{t} &= \PM{s}{t-1} - |\alpha_v[\mb{2}{}]|, \nonumber\\
\PM{3}{t} &= \PM{s}{t-1} - |\alpha_v[\mb{1}{}]| - |\alpha_v[\mb{2}{}]|; \nonumber
\end{align}
where $\PM{0}{t}$ corresponds to the ML decision, $\PM{1}{t}$ to the ML decision with the least-reliable bit flipped, $\PM{2}{t}$ to the ML decision with the second least-reliable bit flipped, and $\PM{3}{t}$ to the ML decision with the two least-reliable bits flipped.

\subsection{SPC Decoders}
\label{sec:algo:spc}
Codewords of an SPC code must satisfy the even parity constraint, i.e. $\sum_i \beta_j[i] = 0$ where the summation is performed using binary arithmetic. As such, $2^{N_v - 1}$ candidate codewords are available, leading to impractical implementations with exponential complexity. Similar to the rate-1 decoders, we use the Chase-II candidate generation to limit the number of candidates. Simulation results, presented in Section~\ref{sec:perf}, showed that flipping combinations of the four least-reliable bits caused only a minor degradation in error-correction performance for $L < 16$ and SPC code length $> 4$. The error-correction performance change was negligible for smaller $L$ values. Increasing the number of least-reliable bits under consideration decreased the decoder speed to the point where not utilizing specialized decoders for SPC codes of length $> 4$ yielded a faster decoder.

We define $q$ as an indicator function so that $q = 1$ when the parity check is satisfied and 0 otherwise. Using this notation, the reliabilities of the candidates, in an expanded form of \eqref{eq:pm}, are
\begin{align}
\PM{0}{t} &= \PM{s}{t-1} - (1-q) \ma{1} \nonumber\\
\PM{1}{t} &= \PM{0}{t} - q \ma{1} - \ma{2},\nonumber\\
\PM{2}{t} &= \PM{0}{t} - q \ma{1} - \ma{3},\nonumber\\
\PM{3}{t} &= \PM{0}{t} - q \ma{1} - \ma{4},\nonumber\\
\PM{4}{t} &= \PM{0}{t} - \ma{2} - \ma{3},\nonumber\\
\PM{5}{t} &= \PM{0}{t} - \ma{2} - \ma{4},\nonumber\\
\PM{6}{t} &= \PM{0}{t} - \ma{3} - \ma{4},\nonumber\\
\PM{7}{t} &= \PM{0}{t} - q \ma{1} \nonumber\\
          &- \ma{2} - \ma{3} - \ma{4};\nonumber
\end{align}
where $\PM{0}{t}$ is reliability of the ML decision calculated according to \eqref{eq:spc}.
The remaining reliability values correspond to flipping an even number of bits compared to the ML decision so that the single-parity check constraint remains satisfied. Applying this rule when the input already satisfies the SPC constraints generates candidates where no bits are flipped, two bits are flipped, and four bits are flipped. Otherwise, one and three bits are flipped.

When the list size $L = 2$, at most two candidates from any given source path are retained. Therefore, only the two most reliable candidates, corresponding to $\PM{0}{t}$ and $\PM{1}{t}$, need to be evaluated for each each source path, regardless of the length of the SPC code. This is supported by the simulation results shown in Section~\ref{sec:perf}.

\subsection{Repetition Decoders}
\label{sec:algo:rep}
A repetition decoder has two possible outputs: the all-zero and the all-one codewords and, according to \eqref{eq:pm}, their reliabilities are
\begin{align}
\PM{0}{t} & = \PM{s}{t-1} - \sum_i h(\alpha_v[i])|\alpha_v[i]|, \nonumber\\
\PM{1}{t} & = \PM{s}{t-1} - \sum_i \left|1 - h(\alpha_v[i])\right||\alpha_v[i]|. \nonumber\\
& = \PM{s}{t-1} - \sum_i h(-\alpha_v[i])|\alpha_v[i]|. \nonumber
\end{align}
where $\PM{0}{t}$ and $\PM{1}{t}$ are the path reliability values corresponding to the all-zero and all-one codewords, respectively. The all-zero reliability is penalized for every input corresponding to a `1' estimate, i.e. negative LLR; and the all-one for every input corresponding to a `0' estimate. These two equations can be rewritten as
\begin{align}
\PM{0}{t} & = \PM{s}{t-1} - \sum_i |\min(\alpha_v[i], 0)|, \nonumber\\
\PM{1}{t} & = \PM{s}{t-1} - \sum_i |\max(\alpha_v[i], 0)|; \nonumber
\end{align}

The ML decision is found according to $\argmax_i (\PM{i}{t})$, which is the same as performing \eqref{eq:rep}.

\section{Implementation}
\label{sec:impl}
In this section we describe the methods used to implement our proposed algorithm on an x86 CPU supporting SIMD instructions. We created two versions: one for CPUs that support the AVX instructions, and the other using SSE for CPUs that do not. For brevity, we only discuss the AVX implementation when both implementations are similar. In cases where they differ significantly, both implementations are presented.

We use 32-bit floating-point (\texttt{float}) to represent the binary-valued $\beta$, in addition to the real-valued $\alpha$, since it improves vectorization of the $g$ operation as explained in Section~\ref{sec:impl:r0rr}.

\subsection{Memory Layout for $\alpha$ Values}
\label{sec:impl:memory}
Memory is organized into stages: the input to all constituent codes of length $N_v$ is stored in stage $S_{\log_2 N_v}$. Due to the sequential nature of the decoding process, only $N_v$ values need to be stored for a stage since old values are discarded when new ones are available. For example, the input to SPC node of size 4 in Fig.~\ref{fig:fast-ssc}, will be stored in $S_2$, overwriting the input to the repetition node of the same size.

When using SIMD instructions, memory must be aligned according the SIMD vector size: 16-byte and 32-byte boundaries for SSE and AVX, respectively. In addition, each stage is padded to ensure that its size is at least that of the SIMD vector. Therefore, a stage of size $N_v$ is allocated $\max(N_v, V)$ elements, where $V$ is the number of $\alpha$ values in a SIMD vector, and the total memory allocated for storing $\alpha$ values is
\[
N + L \sum_{i = 0} ^ {\log_2 N - 1} \max(2^i, V)
\]
LLR (\texttt{float}) elements; where the values in stage $S_{\log_2 N}$ are the channel reliability information that are shared among all paths and $L$ is the list size.

During the candidate forking process at a stage $S_i$, a path $p$ is created from a source path $s$. The new path $p$ shares all the information with $s$ for stages $\in [S_{\log N}, S_i)$. This is exploited in order to minimize the number of memory copy operations by updating memory pointers when a new path is created \cite{Tal2015}. For stages $\in [S_0, S_i]$, path $p$ gets its own memory since the values stored in these stages will differ from those calculated by other descendants of $s$.

\subsection{Memory Layout for $\beta$ Values}
Memory for $\beta$ values is also arranged into stages. However, since calculating $\beta_v$ \eqref{eq:combine} requires both $\beta_l$ and $\beta_r$, values from left and right children are stored separately and do not overwrite each other. Once alignment and padding are accounted for, the total memory required to store $\beta$ values is
\[
L * (N + 2 \sum_{i = 0}^{\log_2N - 1} \max(2^i, V)).
\]
As stage $S_{\log N}$ stores the output candidate codewords of the decoder, which will not be combined with other values, only $L$, instead of $2L$, memory blocks are required.

Stored $\beta$ information is also shared by means of memory pointers. Candidates generated at a stage $S_i$ share all information for stages $\in [S_0, S_i)$.

\subsection{Rate-$R$ and Rate-0 Nodes}
\label{sec:impl:r0rr}
Exploiting the sign-magnitude floating-point representation defined in IEEE-754, allows for efficient vectorized implementation of the $f$ operation \eqref{eq:f}. Extracting the sign and calculating the absolute values in \eqref{eq:f} become simple bit-wise \texttt{AND} operations with the appropriate mask.

The $g$ operation can be written as
\begin{align}
&g(\alpha_v[i], \alpha_v[i + N_v/2], \beta_l[i])\nonumber\\
=& \alpha_v[i + N_v/2] + \beta_l[i] * \alpha_v[i].\nonumber
\end{align}
If we use $\beta \in \{+0.0, -0.0\}$ instead of $\{0, 1\}$, the $g$ operation \eqref{eq:g} can be implemented as
\begin{equation}
\label{eq:opt-g}
\alpha_v[i + N_v/2] + \beta_l[i] \oplus \alpha_v[i].
\end{equation}
Replacing the multiplication ($*$) with an XOR ($\oplus$) operation in \eqref{eq:opt-g} is possible due to the sign-magnitude representation of IEEE-754.

Listing~\ref{lst:f-g} shows the corresponding AVX implementations, originally presented in \cite{Sarkis2014a,Giard2015}, of the $f$ and $g$ functions using the SIMD intrinsic functions provided by GCC. For clarity of exposition, m256 is used instead of \_\_m256 and the \_mm256\_ prefix is removed from the intrinsic function names.

{
\floatname{algorithm}{Listing}
\begin{algorithm}[t]
\caption{Vectorized $f$ and $g$ functions}
\label{lst:f-g}
\begin{lstlisting}[mathescape]
template<unsigned int $N_v$>
void $G$($\alpha^*$ $\alpha_{in}$, $\alpha^*$ $\alpha_{out}$, $\beta^*$ $\beta_{in}$) {
  for (unsigned int i = 0; i < $N_v/2$; i += 8) {
    m256 $\alpha_l$ = load_ps($\alpha_{in}$ + i);
    m256 $\alpha_r$ = load_ps($\alpha_{in}$ + i + $N_v/2$);
    m256 $\beta_l$ = load_ps($\beta_{in}$ + i);
    m256 $\alpha_l'$ = xor_ps($\beta_l$, $\alpha_l$);
    m256 $\alpha_o$ = add_ps($\alpha_r$, $\alpha_l'$);
    store_ps($\alpha_{out}$ + i, $\alpha_o$);
  }
}
template<unsigned int $N_v$>
void $F$($\alpha^*$ $\alpha_{in}$, $\alpha^*$ $\alpha_{out}$) {
  for (unsigned int i = 0; i < $N_v/2$; i += 8) {
    m256 $\alpha_l$ = load_ps($\alpha_{in}$ + i);
    m256 $\alpha_r$ = load_ps($\alpha_{in}$ + i + $N_v/2$);
    m256 sign = and_ps(xor_ps($\alpha_l$, $\alpha_r$), SIGN_MASK);
    m256 |$\alpha_l$| = andnot_ps($\alpha_l$, SIGN_MASK);
    m256 |$\alpha_r$| = andnot_ps($\alpha_r$, SIGN_MASK);
    m256 $\alpha_o$ = or_ps(sign, min_ps(|$\alpha_l$|, |$\alpha_r$|));
    store_ps($\alpha_{out}$ + i, $\alpha_o$);
  }
}
\end{lstlisting}
\end{algorithm}
}

Rate-0 decoders set their output to the all-zero vector using store instructions. The path reliability (PM) calculation \eqref{eq:rel:frozen} is implemented as in Listing~\ref{lst:frozen}.
{
\floatname{algorithm}{Listing}
\begin{algorithm}[t]
\caption{Path reliability update in Rate-0 decoders.}
\label{lst:frozen}
\begin{lstlisting}[mathescape]
m256 ZERO = set1_ps(0.0);
m256 PMv = ZERO;
for (unsigned int i = 0; i < $N_v/2$; i += 8) {
  PMv = add_ps(PMv, min_ps(load_ps($\alpha_{in}$ + i), ZERO));
}
PM = $\sum_i$ PMv$[i]$;
\end{lstlisting}
\end{algorithm}
}

\subsection{Rate-1 Nodes}
\label{sec:impl:info}
Since $\beta \in \{+0.0, -0.0\}$ and $\alpha$ values are represented using sign-magnitude notation, the threshold detection in \eqref{eq:info} is performed using a bit mask (SIGN\_MASK).

Sorting networks can be implemented using SIMD instructions to efficiently sort data on a CPU \cite{Furtak2007}. For rate-1 nodes of length 4, a partial sorting network (PSN), implemented using SSE instructions, is used to find the two least reliable bits. For longer constituent codes, the reliability values are reduced to two SIMD vectors: the first, $v_0$ containing the least reliable bit and the second, $v_1$, containing the least reliable bits not included in $v_0$. When these two vectors are partially sorted using the PSN, $\mb{2}$ will be either the second least-reliable bit in $v_0$ or the least-reliable bit in $v_1$.

\subsection{Repetition Nodes}
\label{sec:impl:rep}
The reliability of the all-zero output $\PM{0}{t}$ is calculated by accumulating the $\min(\alpha_v[i], 0.0)$ using SIMD instructions. Similarly, to calculate $\PM{1}{t}$, $\max(\alpha_v[i], 0.0)$ are accumulated.

\subsection{SPC Nodes}
\label{sec:impl:spc}
For SPC decoders of length 4, all possible bit-flip combinations are tested; therefore, no sorting is performed on the bit reliability values. For longer codes, a sorting network is used to find the four least-reliable bits. When $L = 2$, only the two least reliable bits need to be located. In that case, a partial sorting network is used as described in Section~\ref{sec:impl:info}.

Since the SPC code of length 2 is equivalent to the repetition code of the same length, we only implement the latter.

\section{Adaptive Decoder}
\label{sec:adaptive}
The concatenation with a CRC provides a method to perform early termination analogous to a syndrome check in belief propagation decoders. In \cite{Li2012}, this was used to gradually increase the list size. In this work, we first decode using a Fast-SSC polar decoder, and if the CRC is not satisfied, switch to the list decoder with the target $L_{\max}$ value. The latency of this adaptive approach is
\begin{equation}
\latency{A_{\text{max}}} = \latency{L} + \latency{F};
\end{equation}
where $\latency{L}$ and $\latency{F}$ are the latencies of the list and Fast-SSC decoders, respectively.

The improvement in throughput stems from the Fast-SSC having lower latency than the list decoder. Once the frame-error rate (FER$_{\text{F}}$) at the output of the Fast-SSC decreases below a certain point, the overhead of using that decoder is compensated for by not using the list decoder. The resulting information throughput in bit/s is
\begin{equation}
\label{eq:tp}
\mathcal{T} = \frac{k}{(1 - \text{FER}_{\text{F}})\latency{F} + \text{FER}_{\text{F}}\latency{L}}.
\end{equation}

Determining whether to use adaptive decoder depends on the expected channel conditions and the latency of the list decoder as dictated by $L_{\max}$. This is demonstrated in the comparison with the LDPC codes in Section~\ref{sec:ldpc}.

\section{Performance}
\label{sec:perf}
\subsection{Methodology}
All simulations were run on a single core of an Intel i7-2600 CPU with a base clock frequency of 3.4 GHz and a maximum turbo frequency of 3.8 GHz. Software-defined radio (SDR) applications typically use only one core for decoding, as the other cores are reserved for other signal processing functions \cite{Demel2015}. The decoder was inserted into a digital communication link with binary phase-shift keying (BPSK) and an additive white Gaussian noise (AWGN) channel with random codewords.

Throughput and latency numbers include the time required to copy data to and from the decoder and are measured using the high precision clock from the Boost Chrono library. We report the decoder speed with turbo frequency boost enabled, similar to \cite{Han2013}.

We use the term polar-CRC to denote the result of concatenating a polar code with a CRC. This concatenated code is decoded using a list-CRC decoder. The dimension of the polar code is increased to accommodate the CRC while maintaining the overall code rate; e.g. a (1024, 512) polar-CRC code with an 8-bit CRC uses a (1024, 520) polar code.

\subsection{Choosing a Suitable CRC Length}
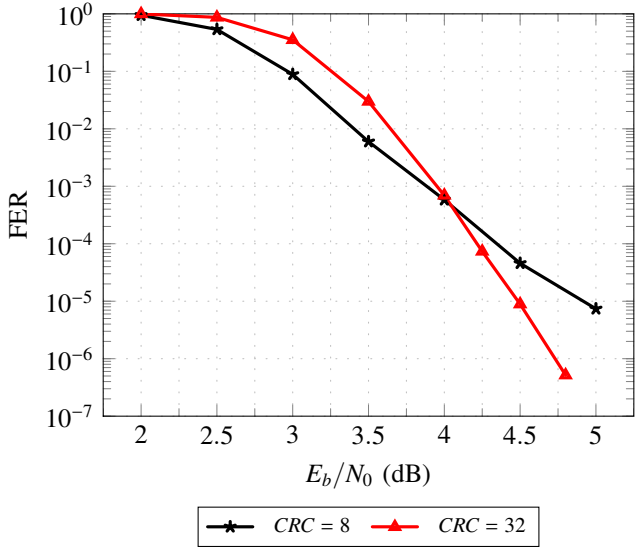
\begin{figure}[t]
 \centering
 \begin{tikzpicture}
  \begin{semilogyaxis}
    [
    width=3.4in,
    height=2.73in,
    line cap=round,
    every axis y label/.style={at={(ticklabel cs:0.5)},rotate=90,anchor=near ticklabel},
    every axis x label/.style={at={(ticklabel cs:0.5)},anchor=near ticklabel},
    xmin=1.75,
    xmax=5.25,
    xlabel={$E_b/N_0$ (dB)},
    minor x tick num={1},
    ymin=1e-7,
    ymax=1,
    ylabel={FER},
    legend columns = 2,
    legend style={
      at={(0.5, -0.275)},
      anchor={center},
      cells={anchor=west},
      column sep= 2mm,
      font=\footnotesize,
    },
    grid=major,
    xminorgrids=true,
    yminorgrids=false,
    grid style=loosely dotted,
    ]

    \addplot[color=black,mark=star, very thick] table[x=snr_db,y=FER] {data/pc1024_860_l128_c08_listcrc.csv};
    \addlegendentry{$CRC=8$}

    \addplot[color=red,mark=triangle, very thick] table[x=snr_db,y=FER] {data/pc1024_860_l128_c32_listcrc.csv};
    \addlegendentry{$CRC=32$}

  \end{semilogyaxis}
\end{tikzpicture}
 \caption{The effect of CRC length on the error-correction performance of $(1024, 860)$ list-CRC decoders with $L=128$.}
 \label{fig:crc_floor}
\end{figure}
Using a CRC as the final output selection criterion significantly improves the error-correction performance of the decoder.
The length of the chosen CRC also affects the error-correction performance depending on the channel conditions. Fig.~\ref{fig:crc_floor} demonstrates this phenomenon for an $(1024, 860)$ polar-CRC code using 8- and 32- bit CRCs and $L=128$. Such a large list size was chosen to ensure that any observed differences are solely due to the change in the CRC length and could not be counteracted by increasing the list size further. The figure shows that the performance is better at lower $E_b/N_0$ values when the shorter CRC is used. The trend is reversed for better channel conditions where the 32-bit CRC provides an improvement $> 0.5$ dB compared to the 8-bit one.

Therefore, the length of the CRC can be selected to improve performance for the target channel conditions.

\subsection{Error-Correction Performance}
\begin{figure}
  \centering
  \begin{tikzpicture}
  \begin{semilogyaxis}
    [
    width=3.4in,
    height=2.73in,
    line cap=round,
    every axis y label/.style={at={(ticklabel cs:0.5)},rotate=90,anchor=near ticklabel},
    every axis x label/.style={at={(ticklabel cs:0.5)},anchor=near ticklabel},
    xmin=3,
    xmax=5,
    xlabel={$E_b/N_0$ (dB)},
    minor x tick num={1},
    ymin=1e-8,
    ymax=1,
    ylabel={FER},
    ytickten={0, -2, -4, -6, -8},
    legend columns = 2,
    legend style={
      at={(0.5, -0.35)},
      anchor={center},
      cells={anchor=west},
      column sep= 2mm,
      font=\footnotesize,
    },
    grid=major,
    xminorgrids=true,
    yminorgrids=false,
    grid style=loosely dotted,
    ]
    
    \addplot[color=orange,very thick, mark=triangle, mark options={solid}] table[x=ebn0_db,y=FER] {data/a2k7-02.dat};
    \addlegendentry{$L = 2$}

    \addplot[color=black,very thick] table[x=ebn0_db,y=FER] {data/10gige.dat};
    \addlegendentry{LDPC (2048,1723)}

    \addplot[color=blue,very thick, mark=square, mark options={solid}] table[x=ebn0_db,y=FER] {data/a2k7-08.dat};
    \addlegendentry{$L = 32$, Dec-SPC-4}

    \addplot[color=blue,very thick, dashed, mark=square, mark options={solid}] table[x=ebn0_db,y=FER] {data/a2k7-11.dat};
    \addlegendentry{$L = 32$, Dec-SPC-$4+$}

    \addplot[color=purple,very thick, mark=o, mark options={solid}] table[x=ebn0_db,y=FER] {data/a2k7-09.dat};
    \addlegendentry{$L = 8$, Dec-SPC-4}

    \addplot[color=purple,very thick,dashed, mark=o, mark options={solid}] table[x=ebn0_db,y=FER] {data/a2k7-10.dat};
    \addlegendentry{$L = 8$, Dec-SPC-$4+$}

  \end{semilogyaxis}

\end{tikzpicture}
  \caption{FER of the polar-CRC (2048, 1723) code using the proposed decoder with different list sizes, with and without SPC decoders.}
  \label{fig:perf}
\end{figure}
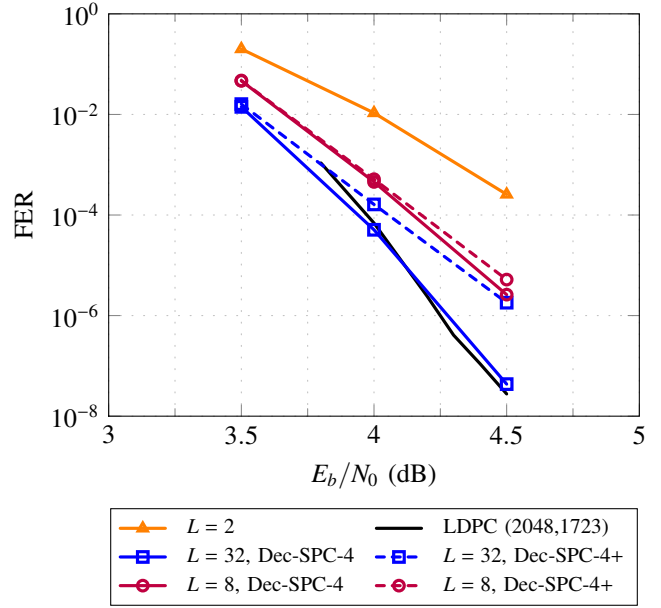

The error-correction performance of the proposed decoder matches that of the SC-List decoder when no SPC constituent decoders of lengths greater than four are used. The longer SPC constituent decoders, denoted SPC-$8+$, only consider the four least-reliable bits in their inputs. This approximation only affects the performance when $L > 2$. Fig.~\ref{fig:perf} illustrates this effect by comparing the FER of different list sizes with and without SPC-$8+$ constituent decoders, labeled Dec-SPC-4 and Dec-SPC-4+, respectively. Since for $L = 2$, the SPC constituent decoders do not affect the error-correction performance, only one graph is shown for that size. As $L$ increases, the FER degradation due to SPC-$8+$ decoders increases. The gap is $< 0.1$ dB for $L = 8$, but grows to $\approx 0.25$ dB when $L$ is increased to 32. These results were obtained with a CRC of length 32 bits. The figure also shows the FER of the (2048, 1723) LDPC code \cite{802.3an} after 10 iterations of offset min-sum decoding for comparison.

While using SPC-$8+$ constituent decoders degrade the error-correction performance for larger $L$ values, they decrease decoding latency as will be shown in the following section. Therefore, the decision regarding whether to employ them or not depends on the target FER and list size.

\subsection{Latency and Throughput}
To determine the latency improvement due to the new algorithm and implementation, 
we compare in Table~\ref{tab:latency} two unrolled decoders with an LLR-based SC-list decoder implemented according to the method described in \cite{BalatsoukasStimming2015}. The first unrolled decoder does not implement any specialized constituent decoders and is labeled ``unrolled SC-list''. While the other, labeled ``unrolled Dec-SPC-4,'' implements all the constituent decoders described in this work, limiting the length of the SPC ones to four.
We observe that unrolling the SC-list decoder decreases decoding latency by more than 50\%. Furthermore, using the rate-0, rate-1, repetition, and SPC-4 constituent decoders decreases the latency to between 63\% ($L = 2$) and 18.9\% ($L = 32$) that of the unrolled SC-list decoder. The speed improvement gained by using the proposed decoding algorithm and implementation compared to SC-list decoding varies between 18.4 and 11.9 times at list sizes of 2 and 32, respectively.\footnote{The gains over an LL-based SC-list decoder are even more significant: such a decoder has a latency of 20.5 ms for $L = 32$, leading the proposed decoder to have 47 times the speed.}
The impact of unrolling the decoder is more evident for smaller list sizes; whereas the new constituent decoders play a more significant role for larger lists.

\begin{table}[t]
  \caption{Latency (in $\mu$s) of decoding the (2048, 1723) polar-CRC code using the proposed method with different list sizes, with and without SPC decoders compared to that of SC-list decoder. Speedups compared to SC-List are shown in brackets}
  \centering
  \begin{tabular}{r c c c c}
    \toprule
    \multirow{2}{*}{Decoder} & \multicolumn{3}{c}{$L$}\\
    \cmidrule{2-4}
                             & 2 & 8 & 32 \\
    \midrule
    SC-List & 558 & 1450 & 5145\\
    Unrolled SC-list & 193 (2.9$\times$) & 564 (2.6$\times$) & 2294 (2.2$\times$) \\
    Unrolled Dec-SPC-4 & 30.4 (18.4$\times$) & 97.5 (14.9$\times$) & 433 (11.9$\times$) \\
    Unrolled Dec-SPC-4+ & 26.3 (21.2$\times$) & 80.2 (18.1$\times$) & N/A \\
    \bottomrule
  \end{tabular}
  \label{tab:latency}
\end{table}

Table~\ref{tab:latency} also shows the latency for the proposed decoder when no restriction on the length of the constituent SPC decoders is present, denoted ``Unrolled Dec-SPC-4+''. We note that enabling these longer constituent decoder decreases latency by 14\% and 18\% for $L = 2$ and 8, respectively. Due to the significant loss in error-correction performance, we do not recommend using the SPC-8+ constituent decoders for $L > 8$ and therefore do not list the latency of such a decoder configuration.

The throughput of the proposed decoder decreases almost linearly with $L$. For $L = 32$ with a latency of 433 $\mu$s, the information throughput is 4.0 Mbps. As mentioned in Section~\ref{sec:adaptive}, throughput can be improved using adaptive decoding where a Fast-SSC decoder is used before the list decoder. The throughput results for this approach are shown for $L = 8$ and $L = 32$ in Table~\ref{tab:throughput}. As $E_b/N_0$ increases, the Fast-SSC succeeds more often and the impact of the list decoder on throughput is decreased, according to ~\eqref{eq:tp}, until it is becomes negligible as can be observed at 4.5 dB where the throughput for both $L = 8$ and 32 is the same.

\begin{table}[t]
  \caption{Information throughput of the proposed adaptive decoder with $L_{\max} = 32$.}
  \centering
  \begin{tabular}{c c c c}
    \toprule
    \multirow{2}{*}{$L$} & \multicolumn{3}{c}{info. T/P (Mbps)}\\
    \cmidrule{2-4}
    & 3.5 dB & 4.0 dB & 4.5 dB\\
    \midrule
    8 & 32.8 & 92.1 & 196 \\
    32 & 8.6 & 33.0 & 196 \\
    \bottomrule
  \end{tabular}
  \label{tab:throughput}
\end{table}

\section{Comparison with LDPC Codes}
\label{sec:ldpc}
\subsection{Comparison with the (2048, 1723) LDPC Code}
\label{sec:perf:10g}
We implemented a scaled min-sum decoder for the (2048, 1723) LDPC code of \cite{802.3an}. To the best of our knowledge, this is the fastest software implementation of decoder for this code. We used early termination and maximum iteration count of 10. To match the error-correction performance at the same code length, an adaptive polar list-CRC decoder with a list size of 32 and a 32-bit CRC was used as shown in Fig.~\ref{fig:perf}.

Table~\ref{tab:tp:3an} presents the results of the speed comparison between the two decoders. It can be observed that the proposed polar decoder has lower latency and higher throughput throughout the entire $E_b/N_0$ range of interest. The throughput advantages widens from seven to 78 times as the channel conditions improve from 3.5 dB to 4.5 dB. The LDPC decoder has three times the latency of the polar list decoder.

\begin{table}[t]
  \caption{Information throughput and latency of the proposed adaptive decoder with $L_{\max} = 32$ compared to the (2048, 1723) LDPC decoder.}
  \centering
  \begin{tabular}{r c c c c}
    \toprule
    \multirow{2}{*}{Decoder}& \multirow{2}{*}{Latency (ms)} & \multicolumn{3}{c}{info. T/P (Mbps)}\\
    \cmidrule{3-5}
    & & 3.5 dB & 4.0 dB & 4.5 dB\\
    \midrule
    LDPC & 1.6 & 1.1 & 2.0 & 2.5\\
    This work & 0.44 & 8.6 & 33.0 & 196 \\
    \bottomrule
  \end{tabular}
  \label{tab:tp:3an}
\end{table}

\subsection{Comparison with the 802.11n LDPC Codes}
\label{sec:perf:wifi}
The fastest software LDPC decoders in literature are those of \cite{Han2013}, which implement decoders for the 802.11n standard \cite{802.11n} using the same Intel Core i7-2600 as this work.

The standard defines three code lengths: 1944, 1296, 648; and four code rates: 1/2, 2/3, 3/4, 5/6. The work in \cite{Han2013} implements decoders for codes of length 1944 and all four rates using a layered offset-min-sum decoding algorithm with five iterations.

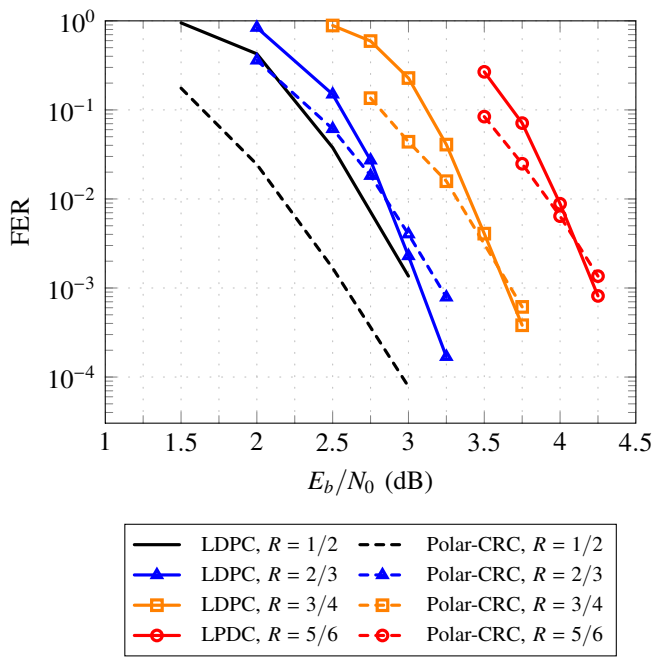
\begin{figure}[t]
  \centering
  \begin{tikzpicture}
  \begin{semilogyaxis}
    [
    width=3.4in,
    height=2.73in,
    line cap=round,
    every axis y label/.style={at={(ticklabel cs:0.5)},rotate=90,anchor=near ticklabel},
    every axis x label/.style={at={(ticklabel cs:0.5)},anchor=near ticklabel},
    xmin=1,
    xmax=4.5,
    ymax=1,
    xlabel={$E_b/N_0$ (dB)},
    minor x tick num={1},
    ylabel={FER},
    legend columns = 2,
    legend style={
      at={(0.5, -0.25)},
      anchor={north},
      cells={anchor=west},
      column sep= 2mm,
      font=\footnotesize,
    },
    grid=major,
    xminorgrids=true,
    yminorgrids=false,
    grid style=loosely dotted,
    ]

    \addplot[color=black,very thick] table[x=ebn0_db,y=FER] {data/wifi5.dat};
    \addlegendentry{LDPC, $R = 1/2$}

    \addplot[color=black,very thick, dashed] table[x=ebn0_db,y=FER] {data/crc8-1k-l2.dat};
    \addlegendentry{Polar-CRC, $R = 1/2$}

    \addplot[color=blue,very thick, mark=triangle, mark options={solid}] table[x=ebn0_db,y=FER] {data/wifi67.dat};
    \addlegendentry{LDPC, $R = 2/3$}

    \addplot[color=blue,very thick,dashed, mark=triangle, mark options={solid}] table[x=ebn0_db,y=FER] {data/crc8-1k-l2-66.dat};
    \addlegendentry{Polar-CRC, $R = 2/3$}

    \addplot[color=orange,very thick, mark=square, mark options={solid}] table[x=ebn0_db,y=FER] {data/wifi75.dat};
    \addlegendentry{LDPC, $R = 3/4$}

    \addplot[color=orange,very thick,dashed, mark=square, mark options={solid}] table[x=ebn0_db,y=FER] {data/crc8-1k-l2-75.dat};
    \addlegendentry{Polar-CRC, $R = 3/4$}

    \addplot[color=red,very thick, mark=o, mark options={solid}] table[x=ebn0_db,y=FER] {data/wifi83.dat};
    \addlegendentry{LPDC, $R = 5/6$}

    \addplot[color=red,very thick,dashed, mark=o, mark options={solid}] table[x=ebn0_db,y=FER] {data/crc8-1k-l2-83.dat};
    \addlegendentry{Polar-CRC, $R = 5/6$}

  \end{semilogyaxis}

\end{tikzpicture}
  \caption{Frame-error rate of the proposed decoders of length 1024 compared with those of the 802.11n standard of length 1944.}
  \label{fig:wifi-perf}
\end{figure}

Fig.~\ref{fig:wifi-perf} shows the FER of these codes using a 10-iteration, flooding-schedule offset min-sum decoder that yields slightly better results than the five iteration layered decoder \cite{Han2013}.
The figure also shows the FER of polar-CRC codes (with 8-bit CRC) of the same rate, but shorter: $N = 1024$ instead of 1944. As can be seen in the figure, when these codes were decoded using a list CRC decoder with $L = 2$, their FER remained within 0.1 dB of the LDPC codes. Specifically, for all codes but the one with rate 2/3, the polar-CRC codes have better FER than their LDPC counterparts down to at least FER $ = 2\times 10 ^{-3}$. For a wireless communication system with retransmission such as 802.11, this constitutes the FER range of interest. These results show that the FER of $N = 1024$ is sufficient and that it is unnecessary to use longer codes to improve it further.

The latency and throughput of the LDPC decoders are calculated for when 524,280 information bits are transferred using multiple LDPC codewords in \cite{Han2013}. Table~\ref{tab:wifi-speed} compares the speed of LDPC and polar-CRC decoders when decoding that many bits on an Intel Core i7-2600 with turbo frequency boost enabled. The latency comprises the total time required to decode all bits in addition to copying them from and to the decoder memory. The results show that the proposed list-CRC decoders are faster than the LDPC ones. The decoder in \cite{Han2013} meets the minimum throughput requirements set in \cite{802.11n} for codes of rate 1/2 and for two out of three cases when the rate is 3/4 (MCS indexes 2 and 3). Our proposed decoder meets the minimum throughput requirements at all code rates.
This shows that in this case, a software polar list decoder obtains higher speeds and similar FER to the LDPC decoder, but with a code about half as long.
Since the decoder operates on individual frames (intra-frame parallelism using SIMD), the latency per frame is significantly lower and is less than 15 $\mu$s for the tested codes as shown in the table.
It should be noted that neither decoder employs early termination: the LDPC decoder in \cite{Han2013} always uses 5 iteration, and the list-CRC decoder does not utilize adaptive decoding.
The number of LDPC and polar code frames required to transmit the 524,280 information bits at each code rate are also shown in Table~\ref{tab:wifi-speed}.

\begin{table}[t]
  \caption{Information throughput and latency of the proposed list decoder compared with the LDPC decoders of \cite{Han2013} when estimating 524,280 information bits.}
  \centering
  \begin{tabular}{r c c c c c c}
    \toprule
    \multirow{2}{*}{Decoder} & \multirow{2}{*}{$N$} & \multirow{3}{*}{\shortstack{\# of $N$-bit\\frames}} & \multirow{2}{*}{Rate} & \multicolumn{2}{c}{Latency (ms)} & \multirow{3}{*}{\shortstack{info. T/P\\(Mbps)}}\\
    \cmidrule{5-6}
    & & & & total & per frame & \\
    \midrule
    \cite{Han2013} & 1944 & 540  & 1/2 & 17.4 & N/A & 30.1\\
    proposed       & 1024 & 1024 & 1/2 & 13.8 & 0.014 & 38.0\\
    \cite{Han2013} & 1944 & 405  & 2/3 & 12.7 & N/A & 41.0\\
    proposed       & 1024 & 768  & 2/3 & 10.0 & 0.013 & 52.4\\
    \cite{Han2013} & 1944 & 360  & 3/4 & 11.2 & N/A & 46.6\\
    proposed       & 1024 & 683  & 3/4 & 8.78 & 0.013& 59.6\\
    \cite{Han2013} & 1944 & 324  & 5/6 & 9.3  & N/A & 56.4\\
    proposed       & 1024 & 615  & 5/6 & 6.2  & 0.010& 84.5\\
    \bottomrule
  \end{tabular}
  \label{tab:wifi-speed}
\end{table}

\section{Conclusion}
In this work, we described an algorithm to significantly reduce the latency of polar list decoding, by an order of magnitude compared to the prior art when implemented in software. We also showed that polar list decoders may be suitable for software-defined radio applications as they can achieve high throughput, especially when using adaptive decoding. Furthermore, when compared with state-of-the art LDPC software decoders from wireless standards, we demonstrated that polar codes could achieve at least the same throughput and similar FER, while using significantly shorter codes. Future work will focus on implementing unrolled list decoders as application-specific integrated circuits (ASIC), which we expect to have throughput approaching 1 Gbps.

\bibliographystyle{IEEEtran}
\bibliography{IEEEabrv,unrolled-list.bib}

\begin{IEEEbiography}[{\includegraphics[width=1in,height=1.25in,clip,keepaspectratio]{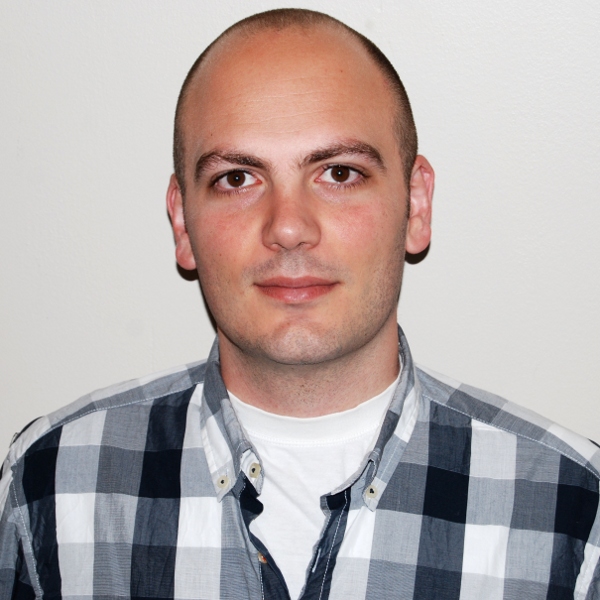}}]{Gabi Sarkis}
Gabi Sarkis received the B.Sc. degree in electrical engineering (summa cum
laude) from Purdue University, West Lafayette, Indiana, United States, in 2006
and the M.Eng. degree from McGill University, Montreal, Quebec, Canada, in
2009. He is currently pursuing a Ph.D. degree at McGill University. His
research interests are in the design of efficient algorithms and
implementations for decoding error-correcting codes, in particular non-binary
LDPC and polar codes.
\end{IEEEbiography}

\begin{IEEEbiography}[{\includegraphics[width=1in,height=1.25in,clip,keepaspectratio]{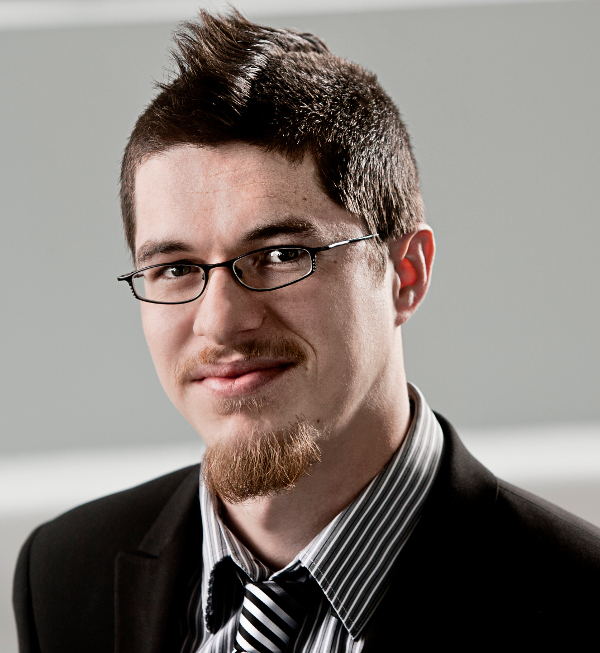}}]{Pascal Giard} received the B.Eng. and M.Eng. degree in
electrical engineering from \'{E}cole de technologie superieure
(\'{E}TS), Montreal, QC, Canada, in 2006 and 2009.
From 2009 to 2010, he worked as a research professional in the
NSERC-Ultra Electronics Chair on 'Wireless Emergency and Tactical
Communication' at \'{E}TS.
He is currently working toward the Ph.D. degree at McGill University.
His research interests are in the design and implementation of signal
processing systems with a focus on modern error-correcting codes.
\end{IEEEbiography}

\begin{IEEEbiography}[{\includegraphics[width=1in,height=1.25in,clip,keepaspectratio]{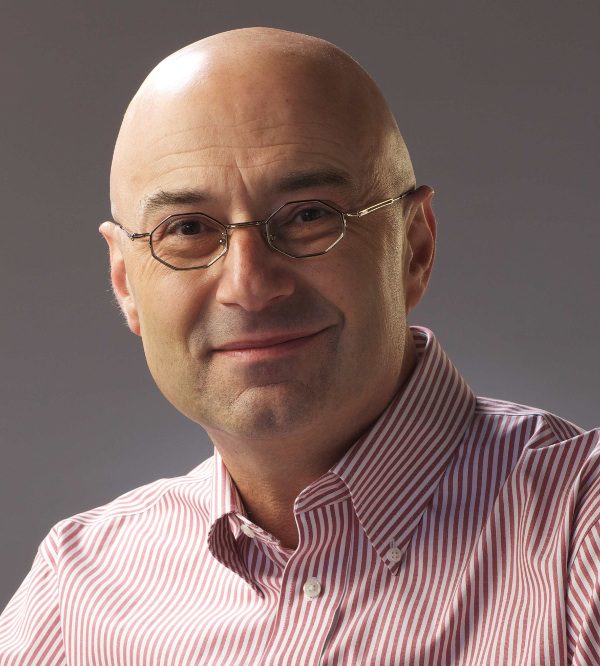}}]{Alexander Vardy}
(S'88--M'91--SM'94--F'99) was born in Moscow, U.S.S.R.,
in 1963. He earned his B.Sc.\ (summa cum laude) from the Technion, Israel,
in 1985, and Ph.D.\ from the Tel-Aviv University, Israel, in 1991. During
1985--1990 he was with the Israeli Air Force, where he worked on electronic
counter measures systems and algorithms. During the years 1992--1993 he
was a Visiting Scientist at the IBM Almaden Research Center, in San Jose, CA.
From 1993 to 1998, he was with the University of Illinois at Urbana-Champaign,
first as an Assistant Professor then as an Associate Professor.
Since 1998, he has been with the University of California San Diego (UCSD),
where he is the Jack Keil Wolf Endowed Chair Professor in the Department
of Electrical and Computer Engineering, with joint appointments in the
Department of Computer Science and the Department of Mathematics.
While on sabbatical from UCSD, he has held long-term visiting appointments
with CNRS, France, the EPFL, Switzerland, and the Technion, Israel.

His research interests include error-correcting codes, algebraic and iterative
decoding algorithms, lattices and sphere packings, coding for digital media,
cryptography and computational complexity theory, and fun math problems.

He received an IBM Invention Achievement Award in 1993, and NSF
Research Initiation and CAREER awards in 1994 and 1995. In 1996, he
was appointed Fellow in the Center for Advanced Study at the University of
Illinois, and received the Xerox Award for faculty research. In the same year,
he became a Fellow of the Packard Foundation. He received the IEEE Information
Theory Society Paper Award (jointly with Ralf Koetter) for the year 2004.
In 2005, he received the Fulbright Senior Scholar Fellowship, and the Best
Paper Award at the IEEE Symposium on Foundations of Computer Science
(FOCS). During 1995--1998, he was an Associate Editor for Coding Theory
and during 1998--2001, he was the Editor-in-Chief of the {\sc IEEE Transactions
on Information Theory}. From 2003 to 2009, he was an Editor for the
SIAM Journal on Discrete Mathematics. He has been a member of the Board
of Governors of the IEEE Information Theory Society during 1998--2006,
and again starting in 2011.
\end{IEEEbiography}

\begin{IEEEbiography}[{\includegraphics[width=1in,height=1.25in,clip,keepaspectratio]{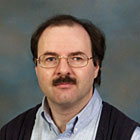}}]{Claude Thibeault}
received his Ph.D. from Ecole Polytechnique de Montreal, Canada. He is now with the Electrical Engineering department of Ecole de technologie superieure, where he serves as full professor.
His research interests include design and verification methodologies targeting ASICs and FPGAs, defect and fault tolerance, as well as current-based IC test and diagnosis. He holds 11 US patents and has published more than 120 journal and conference papers, which were cited more than 550 times. He co-authored the best paper award at DVCON’05, verification category. He has been member of different conference program committee, including the VLSI Test Symposium, for which he was program chair in 2010-2012, and general chair in 2014.
\end{IEEEbiography}

\begin{IEEEbiography}[{\includegraphics[width=1in,height=1.25in,clip,keepaspectratio]{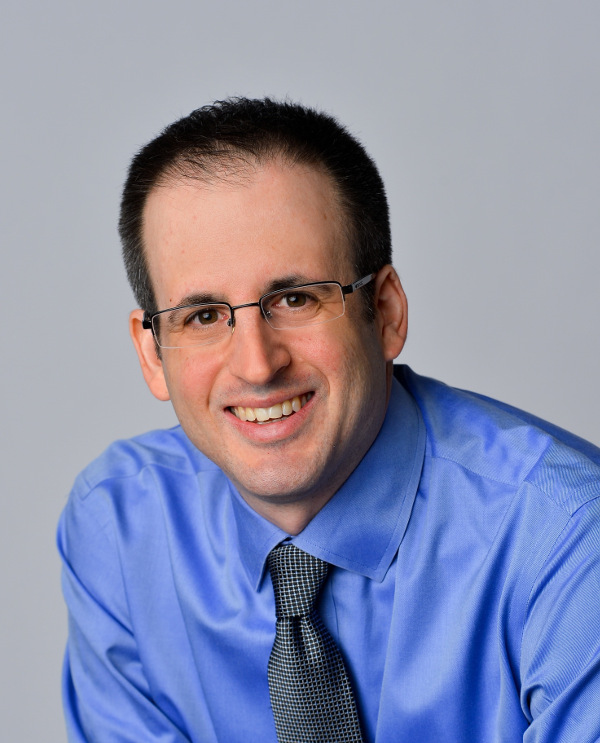}}]{Warren J. Gross}
received the B.A.Sc. degree in electrical engineering from the University of Waterloo, Waterloo, Ontario, Canada, in 1996, and the M.A.Sc. and Ph.D. degrees from the University of Toronto, Toronto, Ontario, Canada, in 1999 and 2003, respectively. Currently, he is an Associate Professor with the Department of Electrical and Computer Engineering, McGill University, Montréal, Québec, Canada. His research interests are in the design and implementation of signal processing systems and custom computer architectures.
Dr. Gross is currently Chair of the IEEE Signal Processing Society Technical Committee on Design and Implementation of Signal Processing Systems. He has served as Technical Program Co-Chair of the IEEE Workshop on Signal Processing Systems (SiPS 2012) and as Chair of the IEEE ICC 2012 Workshop on Emerging Data Storage Technologies. Dr. Gross served as Associate Editor for the IEEE Transactions on Signal Processing. He has served on the Program Committees of the IEEE Workshop on Signal Processing Systems, the IEEE Symposium on Field-Programmable Custom Computing Machines, the International Conference on Field-Programmable Logic and Applications and as the General Chair of the 6th Annual Analog Decoding Workshop. Dr. Gross is a Senior Member of the IEEE and a licensed Professional Engineer in the Province of Ontario.
\end{IEEEbiography}

\end{document}